# Enhancing Medical Image Analysis through Geometric and Photometric transformations

Khadija Rais[1], Mohamed Amroune[1], Mohamed Yassine Haouam[1], Abdelmadjid Benmachiche[2]

[1]Laboratory of mathematics, informatics and systems (LAMIS), Echahid Cheikh Larbi Tebessi University, Tebessa, 12002, Algeria.

[2]Department of Computer Science, LIMA Laboratory, Chadli Bendjedid, University, El-Tarf, PB 73, 36000, Algeria.

E-mails:

Khadija.rais@univ-tebessa.dz; mohamed.amroune@univ-tebessa.dz; mohamed-yassine.haouam@univ-tebessa.dz; benmachiche-abdelmadjid@univ-eltarf.dz;

## Abstract

Medical image analysis suffers from a shortage of labeled data due to several challenges including patient privacy and lack of experts, although some AI models perform well only with large amounts of data and here, we will move to data augmentation where there is a solution to improve the performance of our models and increase the dataset size through traditional or advanced techniques.

In this paper, we evaluate the effectiveness of data augmentation techniques on two different medical image datasets. In the first step, we applied some transformation techniques to the skin cancer dataset containing benign and malignant classes and then trained the convolutional neural network (CNN) on the dataset before and after augmentation, which resulted in a significant improvement in the test accuracy from 90.74% to 96.88% and a decrease in the test loss from 0.7921 to 0.1468 after augmentation. In the second step, we enter Mixup technique by mixing two random images and their corresponding masks using the retina and blood vessel dataset, then we trained the U-net model and obtained the Dice coefficient which was improved from 0 before augmentation to 0.4163 after augmentation.

The result shows the effect of using data augmentation to increase the dataset size on the classification and segmentation performance.

# 1. Introduction

Deep learning has changed medical image analysis by allowing them to create automatic systems with extraordinary accuracy in detecting, classifying, and segmenting various diseases. This propped the innovations for use in clinical diagnosis and decision-making. However, one of the biggest obstacles to applying deep learning to medical imaging is the relative lack of annotated datasets. Medical imaging datasets are typically small due to the challenges associated with data acquisition and labeling, which often require domain expertise and a fair amount of time. This confines the immense capacity for training such robust models that can generalize into unseen data.

Deep learning models succeed greatly when large and diverse datasets are available. Small datasets cause the model to fit well on training data but perform poorly on test data, losing its applicability in real-life scenarios. Data augmentation provides an efficient method of overcoming such cases by artificially increasing the size and variability of datasets [1]. Simple yet effective traditional augmentations such as rotation, flipping, scaling, cropping, and color variations are also used for diversifying training data. They help generalize the model by exposing the network to various transformations of the input data during training and thus increasing its robustness to changes in real-life applications. Mixup is a promising technique in which two random images and their corresponding masks are mixed to create new training samples [2]. This method is smooth and simple compared to deep learning techniques including GAN and VAE [3], resulting in better generalization and robustness, and its benefits have been demonstrated in many different domains, with little impact on medical image analysis, especially segmentation tasks[4].

Our main contribution is to examine the impact of data augmentation techniques on two distinct medical imaging analysis tasks, namely classification and segmentation. We first implement traditional augmentation techniques on a skin cancer dataset containing two classes: benign and malignant. The dataset was distributed to train a Convolutional Neural Network (CNN) to classify skin lesions, and the performance of the model was assessed before augmentations and after augmentations. Secondly, we performed image segmentation by introducing the mixup technique on a retina-blood-vessel dataset using a U-Net model. The need for precise delineation of blood vessels is the main challenge of the segmentation task, and the mixup method introduces novel training samples that help address this challenge.

The results demonstrate the critical role of using data augmentation to improve the performance of medical analysis for both classification and segmentation tasks, with traditional data augmentation techniques demonstrating their improvements using the cancer dataset that requires high accuracy and low loss. Moreover, the mixing technique allows U-Net to perform well and obtain meaningful images for the retinal and vascular dataset. The structure of this paper consists of Section 2 for related work, then methodology in Section 3, discussion and results in Section 4, and finally conclusion in Section 5.

# 2. Related works

Image augmentation increases the dataset size and quality for training by adding more examples to network [5]. Augmentation data techniques are partitioned by researchers into traditional and deep learning-based, where traditional methods involve modifications implemented on data through transformations like flipping, cropping, and noise injection. On the other hand, advanced augmentation methods may incorporate generative models to create entirely new data instances as opposed to mere transformations [6].

In the light of increasing diabetic retinopathy (DR), the necessity of diagnostic system that can cope with time-expensive physician examination and latent lesions (photocoagulation) is addressed, the researchers in [7] The EyePACS dataset was addressed to study the key aspects of such an automated diagnostic solution with focus on data preprocessing, affine transformations and overfitting or class imbalance. Deep learning architecture, which encompasses seven pre-trained deep CADs were evaluated for DR classification. The most performing model of all the architectures trained on EfficientNetV2-M with a test accuracy of 97.65% shows efficacy in this study. Classification metrics like precision, recall, F1 Score, accuracy and loss were utilized to measure the performance. The researchers in another study [8] examined the effect of seven known image augmentation methods on CNN performance in binary classification problems in eleven medical datasets (mainly lung infections and cancer datasets; This included X-ray, USg, PET/CT as well as MRI images). Input augmentation: Vertical and horizontal reflections, fixed random rotations, translations, and crops were tried. All augmentation leads to the creation of one extra copy of every training image in our dataset, thus double dataset. No statistical significance was observed for both US and PET datasets. But Gaussian blur was the best augmentation technique for X-rays and MRI images. The authors in this paper address issues of scarce and imbalanced medical image datasets via Siamese neural networks with approaches like data balancing, weighted loss and constrained augmentation. We achieve up to 5.6% F1 score gain over CNNs on various datasets for COVID-19 diagnosis in the experiments using chest X-ray datasets. Safe augmentations (restricted shifts, scaling and rotations) are employed to keep essential information and increase the robustness of the model, while avoiding any changes that may impact diagnostics [9].

LCAMix is a new technique for data augmentation in medical image segmentation that considers mixing images and masks with contour-aware, superpixel-based techniques. To allow for increased spatial awareness, this will introduce two auxiliary tasks: the classification of local superpixels and the reconstruction of source images. LCAMix is model agnostic, straightforward to implement, and does not need external data; it has shown to outperform various datasets [10]. Deep learning techniques have succeeded in segmenting organ systems but do encounter challenges in lesion segmentation due to data deficiency, morphology diversity, and a lack of informative features. To tackle the above challenges, this paper presents Self-adaptive Data Augmentation (SelfMix), an innovative way towards better lesion segmentation via the self-adaptive fusion of lesion and non-lesion information. Modelled as a unique process unlike existing techniques like Mixup, CutMix, and CarveMix, it contains three major standpoints: (1) less distortion introduced since both tumor and non-tumor information are injected; (2) inclusion into the formula of fusion weights adapted to the lesion geometry and size; (3) brings into consideration non-tumor information. Through experiments on two public datasets, it was shown that SelfMix considerably outperform existing methods in lesion segmentation accuracy [11]. This study investigates whether the mixup data augmentation technique originally designed for classification tasks can enhance the performance of deep segmentation networks in medical imaging. The researchers compared results of U-Net trained on 100 3D T2-weighted MRI scans with and without mixup for prostate segmentation. Metrics such as the Dice similarity coefficient and mean surface distance were used to assess performance. This mixup statistically yielded improvements of up to 1.9% Dice and 10.9% surface distance reduction compared to normal training, with the suggestion that it could be useful in alleviating the issue of data deficit in medical image segmentation [2]. The lack of data to train complex models was addressed by developing a novel dataset extracted from chest CT slices, with the Mixup data augmentation method integrated into a semi-supervised learning frame called Mixup-Inf-Net. Since this procedure utilizes the minimum amount of annotated data and also takes advantage of unlabeled-DATAs over Mixup, the identification of COVID-19-infected areas is possible. Testing on SemiSeg datasets with 3D CT images showed that Mixup-Inf-Net surpasses most of the state-of-the-art segmentation models and is an enhancement on performance learning [12]. This study examines the effect of the mixup data augmentation technique on enhancing the segmentation performance of the U-Net model. The dataset of histopathological images was classified into three groups: (1) augmented with traditional methods (flipping, rotation) images, (2) those augmented only using the mixup method, and (3) images augmented using both techniques. The results indicate that combining mixup with traditional augmentation methods provides an increase in the average Dice coefficient of the model for artifact segmentation [13]. Zhang et al. describe CarveMix, a data augmentation technique used to facilitate convolutional neural network (CNN)-based brain lesion segmentation, which focuses on preserving lesion information through harmonization steps for heterogeneous data and models the mass effect unique to whole brain tumor segmentation. CarveMix stochastically mixes two annotated brain lesion images by carving a ROI based on lesion location and geometry and replacing corresponding voxels in the second image. The results on the multiple datasets indicate that CarveMix successfully enhances the segmentation accuracy of brain lesions [14]. A new augmentation method called TensorMixup uses the three-dimensional U-Net architecture for brain tumor segmentation. The procedure involves selecting patches of image MRI data from two patients using the same modality and mixing them with a tensor that is distributed from a sample of Beta distribution while creating a new image and its corresponding one-hot encoded label. The model trained on this augmented data achieves good segmentation performance, achieving high Dice scores of 92.15%, 86.71%, and 83.49% used for whole tumor, tumor core, and enhancing tumor segmentation, respectively, demonstrating the efficacy of TensorMixup [15].

## 3. Methodology
## 3.1. Data augmentation

Our work aims to improve medical image analysis through data augmentation. Geometric transformation including rotation, scaling, translation, shearing, reflection and optical transformations including brightness adjustment, contrast adjustment, Gaussian noise and histogram equalization are applied to a dataset of benign skin images to generate various variables, as shown in Figure 1: Samples of geometric and photographic transformation. The augmented images are stored for further training of the model.

**Geometric Transformations:** The following transformations were applied to each image:

- **Rotation:** Images rotated 90 degrees clockwise.
- **Scaling:** Images were resized with a scaling factor of 1.2 to simulate zoom effects.
- **Translation:** Images were shifted by 20 pixels horizontally and 30 pixels vertically.
- **Shearing:** A shear transformation matrix was applied to create skew effects.
- **Flipping:** Images were horizontally flipped to create mirrored versions.

**Photometric Transformations:** The following intensity-based transformations were performed to simulate variations in lighting and contrast:

- **Brightness Adjustment:** Brightness levels were increased using pixel value adjustments.
- **Contrast Adjustment**: Contrast levels were amplified using scaling factors.
- **Gaussian Noise Addition:** Random Gaussian noise was added to simulate imaging artifacts.
- **Histogram Equalization:** The luminance channel of the images was equalized to enhance contrast while preserving color fidelity.

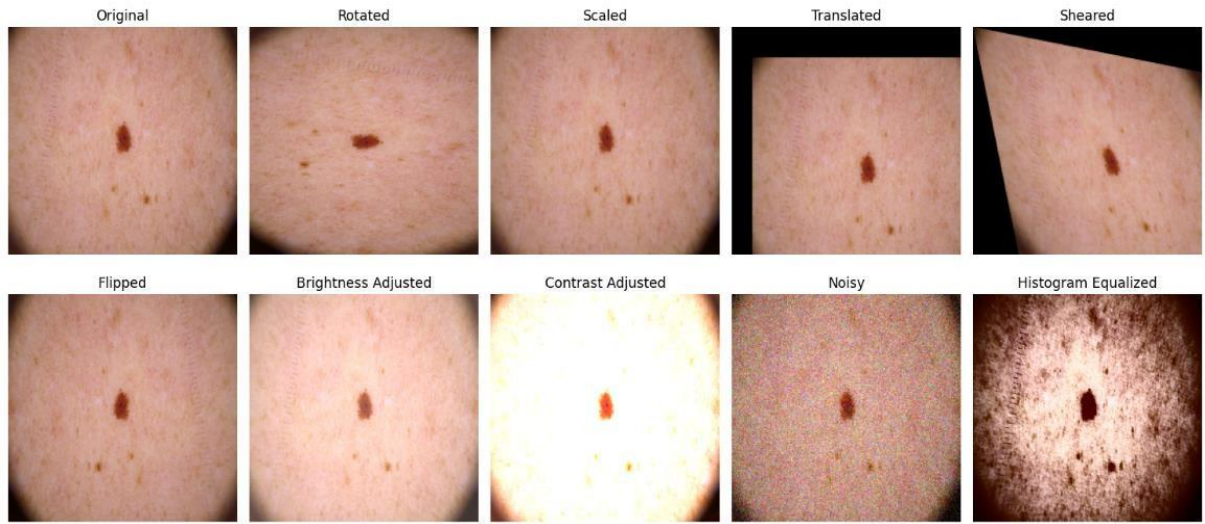

**Figure 1:** Samples of geometric and photographic transformation

**Mixup:** To simulate realistic lesions and interactions between images, a Mixup strategy was implemented. This method combined two images and their corresponding masks to generate new synthetic samples, as shown in Figure 2, preserving lesion information.

Two randomly selected images, along with their masks, were processed:

- Lesion areas were extracted from the first image using the corresponding binary mask.
- Background areas were extracted from the second image.
- A Mixup coefficient (λ) was sampled from a Beta distribution ($\alpha=0.4$) controlling the blend ratio.
- A weighted combination of lesion and background areas was computed for both images and masks, creating mixed images and masks.

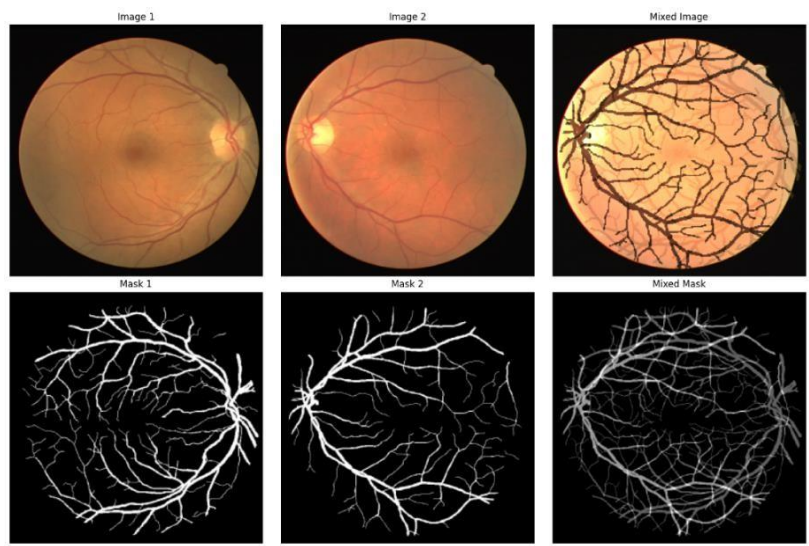

**Figure 2:** Mixup of two images

## 3.2. Dataset

In this study, we used two Kaggle datasets. The first dataset contains skin cancer images [16], which were curated to support tasks such as image classification. They were classified as malignant using 240 images and benign using 30 images. This class was expanded to achieve balance, resulting in 240 images to support accurate discrimination between harmful and harmless parasites. The second dataset shows blood vessels [17] in retinal fundus images, focuses on the segmentation task, contains 100 images and 100 masks, and is divided into 80% images for training and 20% for testing, including their corresponding masks. To enrich this training set, an additional 100 images were augmented.

## 4. Results and discussion

The results for the skin lesion classification task demonstrate the significant impact of data augmentation on CNN performance. Before applying data augmentation, the model achieved an accuracy of 90.74% with a loss of 0.7921. This performance was limited by the severe class imbalance, with only 30 benign images compared to 240 malignant images.

After balancing the dataset through data augmentation, the model's accuracy improved to 96.88%, and the loss decreased to 0.1468, indicating better generalization and robustness. This underscores the importance of data augmentation in handling imbalanced datasets and enhancing the reliability of medical image classification models, Figure 5 presents the losses and accuracies for test data before adding augmented data into train data and Figure 6 presents the losses and accuracies for test data after data augmentation

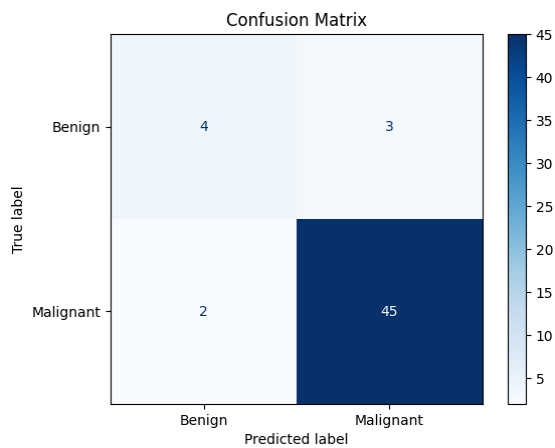

**Figure 3:** Confusion matrix before data augmentation

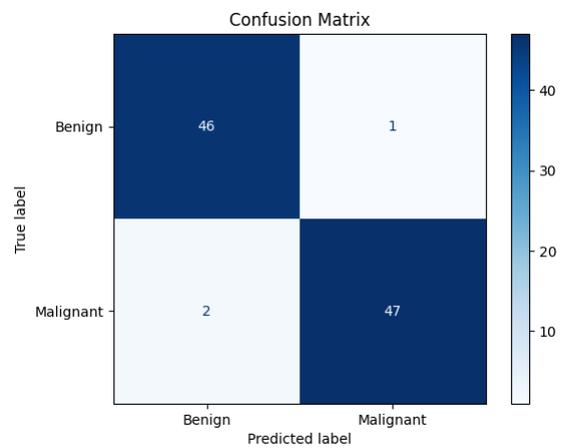

**Figure 4:** Confusion matrix with data augmentation

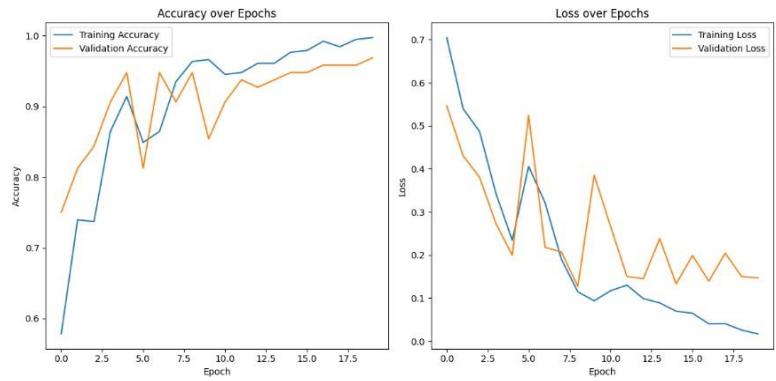

**Figure 5:** training history before data augmentation

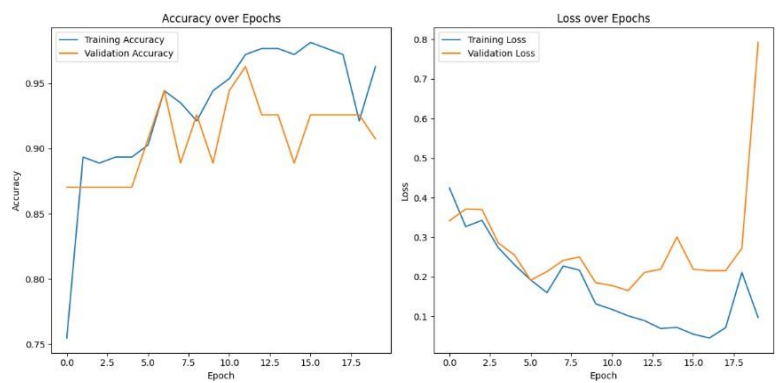

**Figure 6:** training history with data augmentation

For the segmentation task on retinal blood vessels, the dataset initially contained 80 training images and 20 test images with corresponding masks. Before applying data augmentation, the model failed to achieve meaningful segmentation, resulting in a Dice coefficient of 0.00.

By augmenting the training set to include 180 images, the Dice coefficient improved to 0.4163, as mentioned in Table 1, showing a marked improvement in segmentation performance. The additional data likely helped the model learn more diverse patterns, leading to better vessel detection, Figure 7 presented some predicted masks and the dice range from 0.3370 to 0.5887.

The results highlight the critical role of data augmentation in medical image analysis, particularly for datasets with limited or imbalanced samples. For classification, data augmentation not only corrected the imbalance but also enhanced the model's ability to distinguish between benign and malignant lesions. Similarly, in segmentation, augmenting the dataset introduced variability that helped the model achieve more accurate blood vessel segmentation.

However, despite improvements, the Dice score for segmentation remains modest, suggesting room for further enhancement. Techniques like advanced augmentation strategies, better model architectures, or fine-tuning hyperparameters could be explored to improve performance further.

These findings emphasize the necessity of data augmentation in medical image tasks to improve model accuracy and reliability, ultimately aiding in early diagnosis and treatment planning.

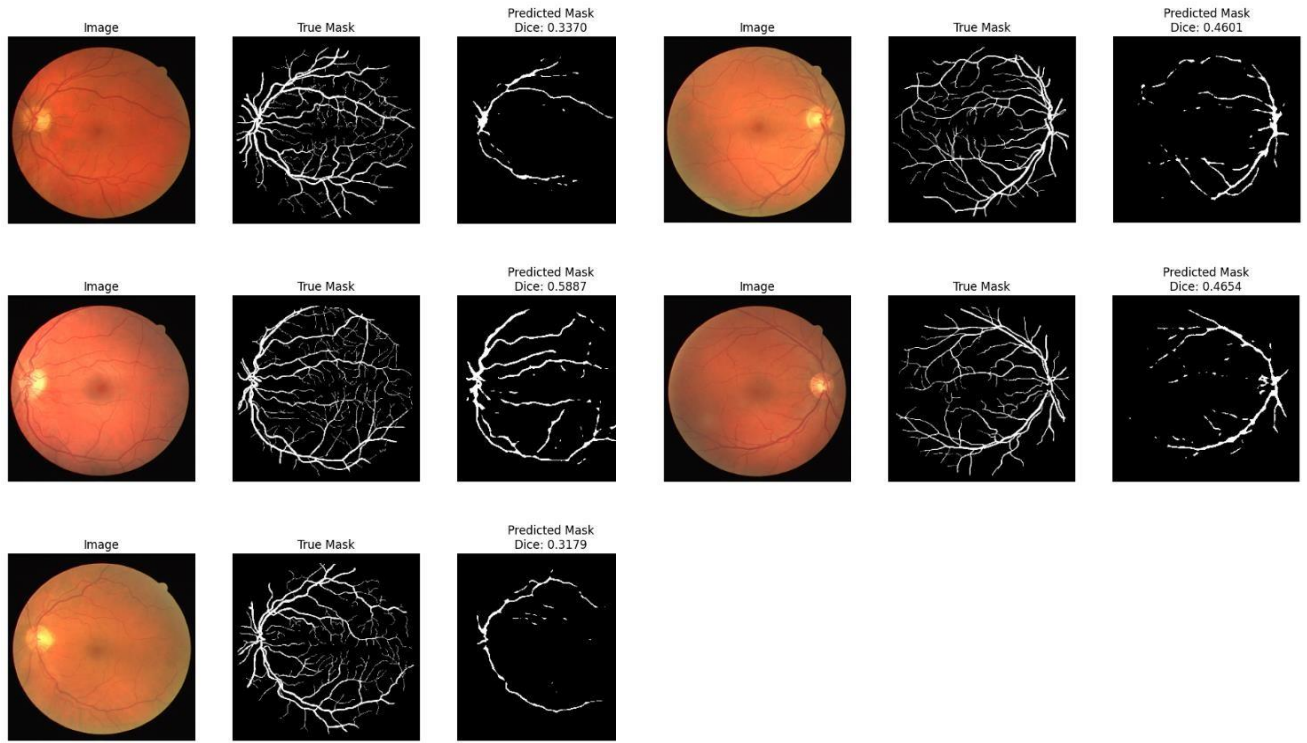

**Figure 7:** Predicted masks

**Table 1:** Classification and segmentation task with and without data augmentation

| Task | Data augmentation (DA) | Dataset | No. of images | | | Metrics | |
|---|---|---|---|---|---|---|---|
| Classification with CNN | Before DA | Skin cancer | Benign | | 30 | Accuracy : 90.74% | Loss : 0.7921 |
| | | | Malignant | | 240 | | |
| | With DA | | Benign | | 240 | Accuracy : 96.88% | Loss : 0.1468 |
| | | | Malignant | | 240 | | |
| Segmentation with U-Net | Before DA | Blood vessels in retinal | Train | Mask | Images | Mean Dice: 0.00 | |
| | | | | 80 | 80 | | |
| | | | Test | Mask | Images | | |
| | | | | 20 | 20 | | |
| | With DA | | Train | Mask | Images | Mean Dice: 0.4163 | |
| | | | | 180 | 180 | | |
| | | | Test | Mask | Images | | |
| | | | | 20 | 20 | | |

## 5. Conclusion

This work underlines the essential role for data augmentation in improving medical image analysis performance, when it is limited or imbalanced datasets that we are dealing with. In the case of skin cancer classification, data augmentation successfully dealt with class imbalance that achieved an accuracy of 96.88% up from 90.74% and significantly reduced the loss. For Retinal Blood Vessel Segmentation, data augmentation on the dataset size and variability lifted making Step from 0.00 to a notable Dice coefficient win of 0.4163.

These results indicate data augmentation can potentially help to ease model overfitting, providing more robust and accurate results to difficult medical image tasks. Ideally, we wish to consider future directions with more elaborate augmentations and architecture, such as performance enhancement or dataset diversity. The current study once again confirms the significance of data augmentation as a pre-processing step to generate robust AI models for medical image analysis and contribute early detection and improved diagnostic outcomes.